\documentclass[12pt,draftclsnofoot,onecolumn]{IEEEtran}

\usepackage[T1]{fontenc}% optional T1 font encoding

\ifCLASSINFOpdf
\else
\fi

\usepackage{amsmath}

\usepackage[cmintegrals]{newtxmath}

\usepackage{graphicx}
\usepackage{subfigure}
\usepackage{multirow}
\usepackage{color}
\usepackage{enumerate}
\usepackage{framed}
\usepackage{mathtools}

\begin{document}

\title{Exact Decoding Probability of Sparse Random Linear Network Coding for Reliable Multicast}

\author{WenLin~Chen, Fang~Lu ~\IEEEmembership{Member,~IEEE,} and Yan~Dong ~\IEEEmembership{Member,~IEEE}

\thanks{The work was supported by the National Nature Science Foundation of China (91538203).
%(\textit{Corresponding author: Yan Dong})
}
\thanks{W. L. Chen, F. Lu and Y. Dong are with the School of Electronic Information and Communications,
Huazhong University of Science and Technology, Wuhan 430074, China
(e-mail: wenlinchen@hust.edu.cn,lufang@hust.edu.cn; dongyan@hust.edu.cn).}}

\markboth{Draft}
{Submitted paper}

\maketitle

\begin{abstract}
Sparse random linear network coding (SRLNC) used as a class of erasure codes to ensure the reliability of multicast communications has been widely investigated. However, an exact expression for the decoding success probability of SRLNC is still unknown, and existing expressions are either asymptotic or approximate. In this paper, we derive an exact expression for the decoding success probability of SRLNC. The key to achieving this is to propose a criterion that a vector is contained in a subspace. To obtain this criterion, we construct a basis of a subspace, with respect to this basis, the coordinates of a vector are known, based on a maximal linearly independent set of the columns of a matrix. The exactness and the computation of the derived expression are demonstrated by a simple example.
\end{abstract}


\IEEEpeerreviewmaketitle

\section{Introduction}\label{section1}
The emerging use cases motivate the design of ultra-reliable and low-latency multicast communications. For example, in automated driving use case, safety-of-life information must reach its destinations with ultra-high reliability within an exceedingly short time frame \cite{5G-PPP}. Traditionally, the reliability of multicast communications is provided by a digital fountain approach \cite{2002Byers}, which is typically realized by LT codes \cite{2002Luby} or Raptor codes \cite{2006Shokrollahi}. However, as noted in \cite{2013Magli}, these kinds of codes achieves its optimal performance only if the number of source packets per generation (called \textit{the generation size}) is large, which leads to large delay.

As an alternative to traditional fountain codes, Sparse Random Linear Network Coding (SRL\\NC), used as a class of erasure codes to ensure reliability of multicast communications, has attracted a lot of interest \cite{2016Tassi,2018Brown}. The SRLNC scheme was originally proposed by Wang \textit{et al.} \cite{2006Wang} to reduce the complexity of Random Linear Network Coding (RLNC), and has been a potential alternative to RLNC. In contrast with traditional fountain codes, SRLNC does not require a large generation size to achieve its optimal performance, which is practically attractive.

In SRLNC for multicast, the source node splits data into the generations, each of which consists of $n$ source packets. During each generation, the source node multicasts coded packets, which are obtained by linearly combining all source packets and using the coefficients selected sparsely from a finite field $F_q$ of order $q$. A destination node recovers $n$ source packets as soon as it collects $n$ linearly independent coded packets. In such multicast network, a key performance metric is the probability of a destination node recovering $n$ source packets from a given number of successfully received coded packets, which is referred to as \textit{the decoding success probability}.
As will be clear in the following section, the decoding success probability of SRLNC is actually the full rank probability of a sparse random matrix over $F_q$.

In the last few decades, many works have devoted to analyze the full rank probability of a sparse random matrix over $F_q$, not only in context of communication, but also in context of probabilistic combinatorics.
Several works have studied the nonsingular probability or the rank distribution of a sparse random matrix over $F_q$ under the asymptotic setting.
In \cite{1990Charlap}, Charlap \textit{et al.} proved that the asymptotic nonsingular probability of a random $n\times n$ matrix over $F_q$ is the same as that for uniform entries, provided that the distribution of the entries of the matrix is not concentrated on any proper affine subspace of $F_q$.
In \cite{2001Kahn}, Kahn \textit{et al.} extended this sufficient condition to a sufficient and necessary condition, that is the distribution of the entries of the matrix is not concentrated on any proper affine subfield of $F_q$.
In \cite{2000-10-Cooper}, Cooper proved that, conditioned on the event that the matrix has no zero rows or columns, the asymptotic rank distribution of a sparse random $(n-s)\times n$ matrix over $F_q$ is the same as that for uniform entries, where $s$ is a nonnegative integer.
In \cite{2011-06-Li,2011-07-Li}, under the assumption of an infinite finite field size, Li \textit{et al.} derived upper and lower bounds on the rank distribution of a sparse random $n\times n$ matrix over $F_q$ by introducing the concept of zero pattern of the random matrix.
Despite the fact that these prior works have advanced our knowledge of the limiting behavior of the rank of the random matrix over $F_q$, the asymptotic results lack accuracy in practical scenarios where neither the generation size $n$ nor the finite field size $q$ is very large.

Many works have studied the performance of SRLNC under the non-asymptotic setting.
In \cite{2016Tassi}, Tassi \textit{et al.} provided an upper-bound on the average number of coded packet transmissions needed to recover the source message for SRLNC, based on an Absorbing Markov Chain (AMC) where the states are defined as the defect of the decoding matrix. However, the transition probabilities are built upon the \textit{BKW} bound \cite[Th. 6.3]{1997Blomer}, which is an upper-bound on the probability $p(i,n)$ of an $n$-dimensional random vector being linearly dependent of $i$ other linearly independent random vectors. As is well known, the \textit{BKW} bound is very loose.
Subsequently, Garrido \textit{et al.} in \cite{2017Garrido} characterized the performance of SRLNC in terms of the decoding success probability and the average number of transmissions required to decode a generation, based on an AMC where the states are defined as the combination of the current rank and the number of non-zero columns of the decoding matrix. However, the transition probabilities rely on Monte Carlo simulations. Obviously, a new set of Monte Carlo simulations are required to re-derive the performance model if system parameters are changed. Recently, Zarei \textit{et al.} in \cite{2020Zarei} focused on the partial decoding delay performance of SRLNC, and presented a lower bound on the average number of transmissions required by a receiver to recover a fraction of a generation, based on an AMC where the states are defined as the combination of the current number of received coded packets and the number of non-zero columns of the decoding matrix.
In \cite{2018Brown}, Brown \textit{et al.} proposed an improved version of the critical set\footnote{In \cite{1994Kolchin}, the critical set is defined as follows. A set $S$ of row numbers is called a critical set if the rows indexed by $S$ sum to the zero vector in $F_q$.}, and hence derived a recursive approximation for the probability of a sparse random $n\times m$ matrix over $F_q$ being full row rank.
This improved statistical event significantly mitigates the impact of the correlation between the critical sets. However, such improved statistical events are not strictly independent. This makes their recursive approximation not tight for large $q$ and $p_0$.

In addition to the recursive non-asymptotic analysis mentioned above, non-recursive non-asymptotic analysis of the performance of SRLNC has been also studied in the literature. In \cite{2019Sehat}, based on linear dependency of a matrix, Sehat \textit{et al.} derived an approximation for the probability $P_{m\times n}^{n}$ of a sparse random $m\times n$ matrix being full column rank. Then by using $P_{m\times n}^{n}$, they presented a recursive equation for the probability $P_{m\times n}^{r}$ of a sparse random $m\times n$ matrix having rank $r$ as a function of $P_{m\times n}^{n}$. However, they do not consider the correlation between linear dependencies of a matrix.
As noted in \cite{2018Brown}, this approximation for $P_{m\times n}^{n}$ is very loose.
In \cite{2019Chen}, based on the reduced row echelon form of a full row rank matrix, Chen \textit{et al.} derived an approximation for $p(i,n)$. Then according to the definition of $p(i,n)$, they established an exact expression for $P_{m,n}^{r}$ as a function of $p(i,n)$.
However, they do not consider the correlation between the entries of a random vector contained in a random subspace. This makes their approximation for $p(i,n)$ not tight in some cases.
Furthermore, due to extremely high computational complexity of the established exact expression for $P_{m\times n}^{r}$ as a function of $p(i,n)$, the proposed approximation for $p(i,n)$ is approximated. This further reduces the tightness of their approximation for $P_{m\times n}^{r}$.
Recently, in \cite{2020Chen}, Chen \textit{et al.} pointed out that the problem of characterizing $P_{m\times n}^{r}$ can be decomposed into two subproblems. The first subproblem is the characterization of $p(i,n)$ (or $P_{m\times n}^{n}$), and the second subproblem is the characterization of $P_{m\times n}^{r}$ as a function of $p(i,n)$ (or $P_{m\times n}^{n}$). In that paper, they presented an exact solution to the second subproblem, based on an AMC where the states are defined as the rank of the decoding matrix, and the eigen decomposition of the transition matrix. In contrast with previous works \cite{2019Sehat,2019Chen}, the exact expression for the second subproblem presented in \cite{2020Chen} is closed-form and has lower complexity.

In summary, an exact expression for the decoding success probability of SRLNC is still unknown, and the existing expressions are either asymptotic or approximate. In this paper, we address this issue by providing an exact expression for the decoding success probability of SRLNC. Based on a maximal linearly independent set of the columns of a matrix, we construct a basis of a subspace, with respect to this basis, the coordinates of a vector are known, hence derive a criterion that a vector is contained in a subspace. By exploiting this criterion, we derive an exact expression for the decoding success probability of SRLNC. The exactness and the computation of the derived expression are demonstrated by a simple example.

The rest of the paper is organized as follows.
Section \ref{section2} describes the considered system model.
In Section \ref{section3}, an exact expression for the decoding success probability of SRLNC is derived.
Section \ref{section4} presents some corollaries of the derived expression.
Section \ref{section5} examines the derived expression by an example. Finally, in Section \ref{section6}, we draw our conclusions.

\section{System Model}\label{section2}
We consider a multicast network where a source node transmits data to multiple destination nodes, and assume that each link from the source node to a destination node is a memoryless packet erasure channel. In order to ensure the reliability of multicast communications, the source node transmits data encoded according to the SRLNC scheme.

The source node splits data into the generations, each of which consists of $n$ source packets $\{x_1,x_2,\cdots,x_n\}$. Each source packet $x_j,~j=1,2,\cdots,n$ consists of $L$ elements from $F_q$. During each generation, the source node injects a stream of coded packets $\{y_1,y_2,\cdots,y_N\}$ into the network.
A coded packet $y_k,~k=1,2,\cdots,N$ is defined as
$y_k=\sum_{j=1}^{n}g_{k,j}x_j$,
where $g_{k,j}\in F_q$ is referred to as the coding coefficient, and the vector $(g_{k,1},g_{k,2}\cdots,g_{k,n})$ is referred to as the coding vector.
In the matrix notation, the encoding process can be expressed as
$(y_1^{T},y_2^{T},\cdots,y_N^{T})^{T}=G(x_1^{T},x_2^{T},\cdots,x_n^{T})^{T}$, where $G=(g_{k,j})$ is an $N\times n$ random matrix over $F_q$.
The coding coefficients $g_{k,j}$ are independent and randomly chosen from $F_q$ according to the following probability distribution:
\begin{equation}\label{equation1}
Pr\{g_{k,j}=t\}=
\begin{cases}
p_0, &t=0 \\
\displaystyle\frac{1-p_0}{q-1}, &t\in F_q\setminus \{0\}
\end{cases}
\end{equation}
where $0\le p_0 \le 1$ is referred to as the sparsity of the code.
The RLNC scheme refers to $p_0=1/q$ (i.e., the coding coefficients are uniformly chosen from $F_q$), and the SRLNC scheme is characterized by $p_0>1/q$.

It is worth mentioning that there is the possibility of the source node generating zero coding vector since the coding vector is randomly generated. From a perspective of real implementation, zero coding vector should not be transmitted since it is ineffectual for recovering the source packets.
However, in order to keep the analysis a higher degree of generality, this paper includes the transmission of zero coding vector as in \cite{2016Tassi,2018Brown}.

In this paper, we focus on the perspective of one destination node. Due to packet loss, each destination node receives a subset of transmitted coded packets.
Let $m$ denote the number of coded packets successfully received by a destination node, where $m\ge n$.
The destination node constructs a $m\times n$ decoding matrix $M$ with $m$ successfully received coded packets. Obviously, the matrix $M$ is obtained from $G$ by deleting the rows corresponding to lost coded packets.
The destination node can recover $n$ source packets if and only if the rank of $M$ is equal to $n$.

\section{Analysis of Decoding Success Probability}\label{section3}
\subsection{Notation}
The term ``independent and identically distributed'' is abbreviated as ``i.i.d.'', and the term ``if and only if'' is abbreviated as ``iff''.

If $v$ is a vector, then $v_{i}$ denotes the $i$-th entry of $v$.
If $M$ is a matrix, then $m_{i}$ denotes either the $i$-th row or the $i$-th column of $M$, the distinction will always be understood from the context.
Let $rank(M)$, $M^{T}$, $M^{-1}$ be the rank, the transpose, and the inverse of a matrix $M$, respectively.
Let $I_{i\times i}$ be the $i\times i$ identity matrix, we omit the subscript of $I_{i\times i}$ to simplify the notation when the context is clear.
Let $<v_1,\cdots,v_i>$ be the linear span of a set of vectors $v_1,\cdots,v_i$.
Let $F_{q}^{n\times m}$ be the set of all $n\times m$ matrices over $F_q$.
Let $wt(\cdot)$ be the weight of a vector or a matrix, i.e., the number of non-zero entries of a vector or a matrix.

\subsection{Exact Formulation}
The analysis of the decoding success probability of SRLNC is conducted by analyzing the probability that a $m\times n$ decoding matrix $M$ is full column rank.

We start by presenting a criterion that a vector is contained in a subspace, which is the basis of the analysis presented in this section.
\newtheorem{lemma}{Lemma}
\begin{lemma}\label{lemma1}
Let $H$ be a random $(i+1)\times n$ matrix over $F_q$, whose entries are i.i.d. with (\ref{equation1}), $0\le i\le n-1$, and assume that the first $i$ rows of $H$ are linearly independent. Let $A$ be an $i\times n$ matrix consisting of the first $i$ rows of $H$, i.e., $A=(h_{1}^{T},h_{2}^{T},\cdots,h_{i}^{T})^{T}$.
Then $H$ is not full row rank iff
\begin{equation*}
\Big((h_{i+1}Q)_{i+1},\cdots,(h_{i+1}Q)_{n}\Big)=\Big((h_{i+1}Q)_{1},\cdots,(h_{i+1}Q)_{i}\Big)(A_{1}^{-1}A_{2}),
\end{equation*}
where $Q$ is the product of the elementary matrices interchanging two columns, $A_1$ is an $i\times i$ matrix consisting of a maximal linearly independent set of the columns of $A$, and $A_2$ is an $i\times (n-i)$ matrix consisting of the remaining columns of $A$ except for $A_1$.
\end{lemma}
\begin{IEEEproof}
It is obvious that $H$ is not full row rank iff the rows $h_1,h_2,\cdots,h_{i+1}$ are linearly dependent. Since $h_1,h_2,\cdots,h_{i}$ are linearly independent, $h_{i+1}$ can be uniquely linearly represented by $h_1,h_2,\cdots,h_{i}$, i.e., there is a unique vector $g=(g_1,g_2,\cdots,g_i) \in F_q^{1\times i}$ such that $h_{i+1}=g_{1}h_{1}+g_{2}h_{2}+\cdots+g_{i}h_{i}$,
or in matrix form,
\begin{equation*}
h_{i+1}=gA.
\end{equation*}
Therefore, $H$ is not full row rank iff $h_{i+1}=gA$ or, equivalently, $h_{i+1}$ is contained in the subspace generated by $h_1,h_2,\cdots,h_{i}$.
However, the distribution of $g$ is unknown. Note that the basis of a vector space is not unique, we can construct a new basis such that the distribution of the coordinates of $h_{i+1}$ with respect to this new basis can be obtained.
In the following, we will show that such basis exists.

We first consider a simple case. If the first $i$ columns of $A$ are linearly independent, then the first $i$ columns of $A$ are a maximal linearly independent set of the columns of $A$, and the last $n-i$ columns of $A$ can be expressed as linear combinations of this maximal linearly independent set.
Let $A_1$ be an $i\times i$ matrix consisting of the first $i$ columns of $A$, and $A_2$ be an $i\times (n-i)$ matrix consisting of the last $n-i$ columns of $A$.
Thus, $rank(A_1)=i$ and $A_2=A_{1}S$, where $S \in F_q^{i\times (n-i)}$ is the coefficient matrix that $A_2$ is linearly expressed by $A_1$. Therefore, we have
\begin{equation*}
A=(A_1,A_2)=(A_1,A_1S)=A_1(I,S)
\end{equation*}
or
\begin{equation*}
A_{1}^{-1}A=(I,S).
\end{equation*}
Let $A'=(I,S)$. According to the change of basis, the rows of $A'$ are also a basis of $<h_1,h_2,\cdots,h_{i}>$ (or the rows of $A'$ generate the same subspace as the rows of $A$).
Therefore, we have
\begin{equation*}
h_{i+1}=g'A',
\end{equation*}
where $g'=(g'_1,g'_2,\cdots,g'_i) \in F_q^{1\times i}$ is a vector whose entries are the coordinates of the vector $h_{i+1}$ with respect to the basis consisting of the rows of $A'$.
By substituting $A'=(I,S)$ into the above equation, we have
\begin{equation*}
h_{i+1}=g'A'=g'(I,S)=(g',g'S).
\end{equation*}
From the above equation, the distribution of $g'$ can be easily obtained, and $H$ is not full row rank iff
\begin{align*}
(h_{i+1,1},h_{i+1,2},\cdots,h_{i+1,i})=g'
\end{align*}
and
\begin{align*}
(h_{i+1,i+1},h_{i+1,i+2},\cdots,h_{i+1,n})=g'S.
\end{align*}

We now consider the general case. If the first $i$ columns of $A$ are not linearly independent, we can interchange the columns of $A$ such that the first $i$ columns of $A$ are linearly independent. This is equivalent to $A$ is multiplied by an $n\times n$ invertible matrix $Q$ on the right such that the first $i$ columns of $A$ are linearly independent, i.e., $AQ=(A_1,A_2)$, where $A_1$ is an $i\times i$ matrix consisting of a maximal linearly independent set of the columns of $A$, and $A_2$ is an $i\times (n-i)$ matrix consisting of the remaining columns of $A$ except for $A_1$. It is worth noting that $(A_1,A_2)$ only differs from $A$ in the order of the columns. Thus, $rank(A_1)=i$ and $A_2=A_{1}S$, where $S\in F_q^{i\times (n-i)}$ is the coefficient matrix that $A_2$ is linearly expressed by $A_1$.
Therefore, we have
\begin{equation*}
AQ=(A_1,A_2)=(A_1,A_1S)=A_1(I,S)
\end{equation*}
or
\begin{equation*}
A_{1}^{-1}A=(I,S)Q^{-1}.
\end{equation*}
Let $A'=(I,S)Q^{-1}$.
According to the change of basis, the rows of $A'$ are also a basis of $<h_1,h_2,\cdots,h_{i}>$. Therefore, we have
\begin{equation*}
h_{i+1}=g'A',
\end{equation*}
where $g'=(g'_1,g'_2,\cdots,g'_i) \in F_q^{1\times i}$ is a vector whose entries are the coordinates of the vector $h_{i+1}$ with respect to the basis consisting of the rows of $A'$. By substituting $A'=(I,S)Q^{-1}$ into the above equation, we have
\begin{equation*}
h_{i+1}=g'A'=g'(I,S)Q^{-1}
\end{equation*}
or
\begin{equation*}
h_{i+1}Q=g'(I,S)=(g',g'S).
\end{equation*}
From the above equation, the distribution of $g'$ can be easily obtained (since the entries of $h_{i+1}$ are i.i.d. with (\ref{equation1}) and $Q$ is the product of the elementary matrices interchanging two columns, the entries of the vector $h_{i+1}Q$ are still i.i.d. with (\ref{equation1}), and hence $h_{i+1}Q$ is essentially the same as $h_{i+1}$), and
$H$ is not full row rank iff
\begin{equation*}
\Big((h_{i+1}Q)_{1},(h_{i+1}Q)_{2},\cdots,(h_{i+1}Q)_{i}\Big)=g'
\end{equation*}
and
\begin{equation*}
\Big((h_{i+1}Q)_{i+1},(h_{i+1}Q)_{i+2},\cdots,(h_{i+1}Q)_{n}\Big)=g'S.
\end{equation*}
This completes the proof.
\end{IEEEproof}

Based on Lemma \ref{lemma1}, we now derive an exact expression for the probability $p(i,n)$ that an $n$-dimensional random vector is linearly dependent of $i$ other linearly independent random vector.
\newtheorem{theorem}{Theorem}
\begin{theorem}\label{theorem1}
Let $H$ be a random $(i+1)\times n$ matrix over $F_q$, whose entries are i.i.d. with (\ref{equation1}), $0\le i \le n-1$, and assume that the first $i$ rows of $H$ are linearly independent. Then the probability of $H$ being not full row rank is given by
\begin{align}\label{equation2}
p(i,n)=&\sum_{z\in F_{q}^{1\times i}}
p_{0}^{i-wt(z)}\Big(\frac{1-p_{0}}{q-1}\Big)^{wt(z)}
\sum_{\substack{C\in F_{q}^{i\times n}\\ rank(C)=i}}
\frac{p_{0}^{in-wt(C)}\Big(\frac{1-p_{0}}{q-1}\Big)^{wt(C)}}
{\sum\limits_{\substack{D\in F_{q}^{i \times n}\\rank(D)=i}}
p_{0}^{in-wt(D)}\Big(\frac{1-p_{0}}{q-1}\Big)^{wt(D)}}\nonumber\\
&p_{0}^{n-i-wt(zC_{1}^{-1}C_{2})}\Big(\frac{1-p_{0}}{q-1}\Big)^{wt(zC_{1}^{-1}C_{2})},
\end{align}
where $C_{1}$ is an $i\times i$ matrix consisting of a maximal linearly independent set of the columns of $C$, and $C_{2}$ is an $i\times (n-i)$ matrix consisting of the remaining columns of $C$ except for $C_1$.

\end{theorem}
\begin{IEEEproof}
According to Lemma \ref{lemma1} and the total probability theorem,
\begin{align*}
&p(i,n)\\
=&Pr\{\Big((h_{i+1}Q)_{i+1},\cdots,(h_{i+1}Q)_{n}\Big)=g'S\}\\
=&\sum_{z\in F_{q}^{1\times i}}Pr\{g'=z\}
Pr\{\Big((h_{i+1}Q)_{i+1},\cdots,(h_{i+1}Q)_{n}\Big)=g'S|g'=z\}\\
=&\sum_{z\in F_{q}^{1\times i}}Pr\{g'=z\}
\sum_{\mathclap{\substack{C\in F_{q}^{i\times n}\\ rank(C)=i}}}Pr\{A=C\}
Pr\{\Big((h_{i+1}Q)_{i+1},\cdots,(h_{i+1}Q)_{n}\Big)=g'S|g'=z,A=C\}.
\end{align*}
The probabilities $Pr\{g'=z\}$, $Pr\{A=C\}$ and $Pr\{\Big((h_{i+1}Q)_{i+1},\cdots,(h_{i+1}Q)_{n}\Big)=g'S~|~g'=z,A=C\}$ are calculated as follows:
\begin{align*}
&Pr\{g'=z\}=p_{0}^{i-wt(z)}\Big(\frac{1-p_{0}}{q-1}\Big)^{wt(z)},\\
&Pr\{A=C\}=\frac{p_{0}^{in-wt(C)}\Big(\frac{1-p_{0}}{q-1}\Big)^{wt(C)}}
{\sum\limits_{\substack{D\in F_{q}^{i \times n}\\rank(D)=i}}
p_{0}^{in-wt(D)}\Big(\frac{1-p_{0}}{q-1}\Big)^{wt(D)}},\\
&Pr\{\Big((h_{i+1}Q)_{i+1},\cdots,(h_{i+1}Q)_{n}\Big)=g'S~|~g'=z,A=C\}\\
=&Pr\{\Big((h_{i+1}Q)_{i+1},\cdots,(h_{i+1}Q)_{n}\Big)=zC_{1}^{-1}C_{2}\}\\
=&p_{0}^{n-i-wt(zC_{1}^{-1}C_{2})}\Big(\frac{1-p_{0}}{q-1}\Big)^{wt(zC_{1}^{-1}C_{2})}.
\end{align*}
This completes the proof.
\end{IEEEproof}

At this point, we have derived an exact expression for the probability $p(i,n)$ that an $n$-dimensional random vector is linearly dependent of $i$ other linearly independent random vectors. We now proceed to derive the probability that a $m\times n$ decoding matrix $M$ is full column rank.

\begin{theorem}\label{theorem2}
Let $M$ be a random $m\times n$ matrix over $F_q$, whose entries are i.i.d. with (\ref{equation1}), $m\ge n$. Then the probability of $M$ being full column rank is given by
\begin{align}\label{equation3}
P_{m\times n}^{n}=\prod_{i=0}^{n-1}\Big(1-p(i,m)\Big).
\end{align}
\end{theorem}
\begin{IEEEproof}
Let $M_{i}$ be a $m\times i$ matrix consisting of the first $i$ columns of $M$. It is obvious that $Pr\{rank(M_{0})=0\}=1$.
Since $M$ has rank $n$ iff the first $i$ columns of $M$ are linearly independent, $0\le i\le n$,
\begin{align*}
Pr\{rank(M)=n\}=Pr\{\bigcap_{i=0}^{n}rank(M_{i})=i\}.
\end{align*}
According to the product theorem,
\begin{align*}
&Pr\{rank(M)=n\}\\
=&Pr\{rank(M_{n})=n~|~\bigcap_{i=0}^{n-1}rank(M_{i})=i\}\\
&Pr\{rank(M_{n-1})=n-1~|~\bigcap_{i=0}^{n-2}rank(M_{i})=i\}\\
&\cdots\\
&Pr\{rank(M_{1})=1~|~rank(M_{0})=0\}\\
&Pr\{rank(M_{0})=0\}\\
=&Pr\{rank(M_{n})=n~|~rank(M_{n-1})=n-1\}\\
&Pr\{rank(M_{n-1})=n-1~|~rank(M_{n-2})=n-2\}\\
&\cdots\\
&Pr\{rank(M_{1})=1~|~rank(M_{0})=0\}\\
&Pr\{rank(M_{0})=0\}\\
=&\prod_{i=0}^{n-1}Pr\{rank(M_{i+1})=i+1~|~rank(M_{i})=i\}\\
=&\prod_{i=0}^{n-1}\Big(1-p(i,m)\Big)
\end{align*}
This completes the proof.
\end{IEEEproof}

\section{Discussion}\label{section4}
In this section, we will show that the derived formulas (\ref{equation2}) and (\ref{equation3}) can collapse to the RLNC case, and (\ref{equation2}) can be scaled up to the well-known \textit{BKW} bound, and (\ref{equation3}) is equivalent to the existing expressions.

\newtheorem{corollary}{Corollary}
\begin{corollary}
(\ref{equation2}) and (\ref{equation3}) is a generalization of the RLNC case.
\end{corollary}
\begin{IEEEproof}
Let $U=\{D\in F_{q}^{i\times n}|rank(D)=i\}$.
By substituting $p_0=1/q$ into (\ref{equation2}),
\begin{align*}
&Pr\{g'=z\}=(1/q)^{i},\\
&Pr\{A=C\}=\frac{(1/q)^{in}}{(1/q)^{in}|U|}=\frac{1}{|U|},\\
&Pr\{\Big((h_{i+1}Q)_{i+1},\cdots,(h_{i+1}Q)_{n}\Big)=g'S|g'=z,A=C\}=(1/q)^{n-i}.
\end{align*}
Therefore,
\begin{align*}
p(i,n)=&\sum_{z\in F_{q}^{1\times i}}Pr\{g'=z\}
\sum_{\substack{C\in F_{q}^{i\times n}\\rank(C)=i}}Pr\{A=C\}\\
&Pr\{\Big((h_{i+1}Q)_{i+1},\cdots,(h_{i+1}Q)_{n}\Big)=g'S|g'=z,A=C\}\\
=&\sum_{z\in F_{q}^{1\times i}}\sum_{\substack{C\in F_{q}^{i\times n}\\rank(C)=i}}
\Big(\frac{1}{q}\Big)^{i}\frac{1}{|U|}\Big(\frac{1}{q}\Big)^{n-i}\\
=&\Big(\frac{1}{q}\Big)^{n-i}.
\end{align*}
It is well-known that for RLNC scheme, the probability of a vector being contained in the subspace generated by $i$ linearly independent vectors is $(1/q)^{n-i}$  \cite{2000Cooper,2011Trullols-Cruces}.
This shows that (\ref{equation2}) can collapse to RLNC case when $p_0=1/q$.

By substituting $p_0=1/q$ into (\ref{equation3}), we have $P_{m\times n}^{n}=\prod_{i=0}^{n-1}(1-(1/q)^{m-i})$. It is easy to check that this equation is equivalent to \cite[eq. (7)]{2011Trullols-Cruces}. This shows that (\ref{equation3}) can collapse to RLNC case when $p_0=1/q$.
\end{IEEEproof}

\begin{corollary}
(\ref{equation2}) can be scaled up to the \textit{BKW} bound.
\end{corollary}
\begin{IEEEproof}
From (\ref{equation2}),
\begin{align*}
&Pr\{\Big((h_{i+1}Q)_{i+1},\cdots,(h_{i+1}Q)_{n}\Big)=g'S|g'=z,A=C\}\\
=&p_0^{n-i-wt(zC_{1}^{-1}C_{2})}\Big(\frac{1-p_0}{q-1}\Big)^{wt(zC_{1}^{-1}C_{2})}\\
\le& \Big(max(p_0,\frac{1-p_0}{q-1})\Big)^{n-i-wt(zC_{1}^{-1}C_{2})}
\Big(max(p_0,\frac{1-p_0}{q-1})\Big)^{wt(zC_{1}^{-1}C_{2})}\\
=&\Big(max(p_0,\frac{1-p_0}{q-1})\Big)^{n-i}.
\end{align*}
Therefore,
\begin{align*}
p(i,n)=&\sum_{z\in F_{q}^{1\times i}}Pr\{g'=z\}
\sum_{\substack{C\in F_{q}^{i\times n}\\rank(C)=i}}Pr\{A=C\}\\
&Pr\{\Big((h_{i+1}Q)_{i+1},\cdots,(h_{i+1}Q)_{n}\Big)=g'S|g'=z,A=C\}\\
\le&\Big(max(p_0,\frac{1-p_0}{q-1})\Big)^{n-i}
\sum_{z\in F_{q}^{1\times i}}Pr\{g'=z\}\sum_{\substack{C\in F_{q}^{i\times n}\\rank(C)=i}}Pr\{A=C\}\\
=&\Big(max(p_0,\frac{1-p_0}{q-1})\Big)^{n-i}.
\end{align*}

In the proof of \cite[Th. 6.3]{1997Blomer}, Bl\"omer \textit{et al.} stated that the probability of the vector $h_{i+1}$ being contained in the subspace generated by the vectors $h_1,h_2,\cdots,h_i$ is at most $\Big(max(p_0,\frac{1-p_0}{q-1})\Big)^{n-i}$.
This completes the proof.
\end{IEEEproof}

\begin{corollary}\label{corollary3}
In terms of the full rank probability of a matrix, (\ref{equation3}) is equivalent to \cite[eq. (4)]{2020Chen}. That is,
\begin{align*}
\prod_{i=0}^{n-1}(1-p(i,m))=\Big(\prod_{t=0}^{n-1}(p(t,n)-1)\Big)
\sum_{k=0}^{n}\frac{(p(k,n))^m}{\prod_{t=0,t\neq k}^{n}(p(t,n)-p(k,n))}.
\end{align*}
\end{corollary}
\begin{IEEEproof}
Let $a_t=p(t-1,n),~t=1,2,\cdots,n+1$. Note that $a_t$ deals with $n$-dimensional row vectors. From \cite[eq. (4)]{2020Chen}, the rank distribution of $M$ is given by
\begin{align*}
P_{m\times n}^{r}=\Big(\prod_{t=1}^{r}(a_t-1)\Big)
\sum_{k=1}^{r+1}\frac{(a_k)^m}{\prod_{t=1,t\neq k}^{r+1}(a_t-a_k)},
\end{align*}
where $r=0,1,\cdots,n$.
In particular, when $r=n$, we have
\begin{align*}
P_{m\times n}^{n}=&\Big(\prod_{t=1}^{n}(a_t-1)\Big)
\sum_{k=1}^{n+1}\frac{(a_k)^m}{\prod_{t=1,t\neq k}^{n+1}(a_t-a_k)}.
\end{align*}

\cite[eq. (4)]{2020Chen} is obtained by using Markov chain from a perspective of the rows of a matrix ($n$-dimensional row vectors and $m$ transitions).
Since the rank of a matrix is equal to the rank of its transpose, we can also use Markov chain from a perspective of the columns of that matrix ($m$-dimensional column vectors and $n$ transitions).

Let $b_t=p(t-1,m),~t=1,2,\cdots,m+1$. Note that $b_t$ deals with $m$-dimensional column vectors. Using the same method as \cite[eq. (4)]{2020Chen}, we have
\begin{align*}
P_{m\times n}^{r}=\Big(\prod_{t=1}^{r}(b_t-1)\Big)\sum_{k=1}^{r+1}\frac{(b_k)^{n}}{\prod_{t=1,t\neq k}^{r+1}(b_t-b_k)},
\end{align*}
where $r=0,1,\cdots,m$.
Since $m\ge n$, the rank of $M$ is at most $n$.
In particular, when $r=n$, we have
\begin{align*}
P_{m\times n}^{n}=\Big(\prod_{t=1}^{n}(b_t-1)\Big)\sum_{k=1}^{n+1}\frac{(b_k)^{n}}{\prod_{t=1,t\neq k}^{n+1}(b_t-b_k)}=\prod_{t=1}^{n}(1-b_t).
\end{align*}
The last equality is proved as follows.

From \cite[Lemma 5]{2020Chen}, we have
\begin{align*}
\sum_{k=1}^{r+1}\frac{(x_{k})^{m}}{\prod_{t=1,t\neq k}^{r+1}(x_t-x_k)}
=(-1)^{r}\sum_{\substack{m_1+m_2+\cdots+m_{r+1}=m-r\\m_{i}\ge 0}}x_{1}^{m_1}x_{2}^{m_2}\cdots x_{r+1}^{m_{r+1}}.
\end{align*}
Let $m=n$ and $r=n$. Then,
\begin{align*}
\sum_{k=1}^{n+1}\frac{(x_k)^{n}}{\prod_{t=1,t\neq k}^{n+1}(x_t-x_k)}=(-1)^{n}.
\end{align*}
This completes the proof.
\end{IEEEproof}

\begin{corollary}
In terms of the full rank probability of a matrix, (\ref{equation3}) is equivalent to \cite[eq. (20)]{2019Chen}. That is,
\begin{align}\label{equation4}
\prod_{i=0}^{n-1}\Big(1-p(i,m)\Big)=\prod_{i=0}^{n-1}(1-p(i,n))
\sum_{i_1=0}^{n}p(i_1,n)\sum_{i_2=i_1}^{n}p(i_2,n)\cdots\sum_{i_{m-n}=i_{m-n-1}}^{n}p(i_{m-n},n).
\end{align}
\end{corollary}
\begin{IEEEproof}
From \cite[eq. (20)]{2019Chen}, an exact expression for $P_{m\times n}^{r}$ as a function of $p(i,n)$ is given by
\begin{align*}
P_{m\times n}^{r}=\prod_{i=0}^{r-1}(1-p(i,n))
\sum_{i_1=0}^{r}p(i_1,n)\sum_{i_2=i_1}^{r}p(i_2,n)\cdots\sum_{i_{m-r}=i_{m-r-1}}^{r}p(i_{m-r},n).
\end{align*}
This expression is derived from the fact that the rank of $M$ is equal to $r$ iff there are $r$ rows of $M$ increasing the rank and the remaining $m-r$ rows of $M$ maintain the rank.

According to Corollary \ref{corollary3}, we have
\begin{align*}
\prod_{i=0}^{n-1}(1-p(i,m))=\Big(\prod_{t=0}^{n-1}(p(t,n)-1)\Big)
\sum_{k=0}^{n}\frac{(p(k,n))^{m}}{\prod_{t=0,t\neq k}^{n}(p(t,n)-p(k,n))}.
\end{align*}
From \cite[Corollary 3]{2020Chen}, we have
\begin{align*}
P_{m\times n}^{r}=&\prod_{t=1}^{r}(p(t-1,n)-1)
\sum_{k=1}^{r+1}\frac{(p(k-1,n))^{m}}{\prod_{t=1,t\neq k}^{r+1}(p(t-1,n)-p(k-1,n))}\\
=&\prod_{t=0}^{r-1}(1-p(t,n))\sum_{i_1=0}^{r}p(i_1,n)\sum_{i_2=i_1}^{r}p(i_2,n)\cdots\sum_{i_{m-r}=i_{m-r-1}}^{r}p(i_{m-r},n).
\end{align*}
This completes the proof.
\end{IEEEproof}
\newtheorem{remark}{Remark}
\begin{remark}
Intuitively, the left hand side of (\ref{equation4}) is derived from a perspective of the columns of $M$, while the right hand side of (\ref{equation4}) is derived from a perspective of the rows of $M$.
\end{remark}

\section{Numerical Results}\label{section5}
In this section, we demonstrate the exactness and the computation of the derived formula by an example. Due to the limitation of space, we provide only one example.

\textit{Example 1:} Consider a $3\times 3$ matrix over $F_2$.

From (\ref{equation3}),
\begin{align*}
P_{3\times 3}^{3}=\prod_{i=0}^{2}(1-p(i,3)).
\end{align*}

(i) $i=0$. Then $p(0,n)=p_0^{n}$. Therefore, $p(0,3)=p_0^{3}$.

(ii) $i=1$. According to the conditional probability measure of random matrix $A$ and $S=A_{1}^{-1}A_2$, we can obtain a probability measure of random matrix $S$.
The conditional probability measure of $A$ is
\begin{align*}
&Pr\{A=(0~0~1)\}=\frac{p_0^{2}(1-p_0)}{1-p_0^{3}},\\
&Pr\{A=(0~1~0)\}=\frac{p_0^{2}(1-p_0)}{1-p_0^{3}},\\
&Pr\{A=(0~1~1)\}=\frac{p_0(1-p_0)^{2}}{1-p_0^{3}},\\
&Pr\{A=(1~0~0)\}=\frac{p_0^{2}(1-p_0)}{1-p_0^{3}},\\
&Pr\{A=(1~0~1)\}=\frac{p_0(1-p_0)^{2}}{1-p_0^{3}},\\
&Pr\{A=(1~1~0)\}=\frac{p_0(1-p_0)^{2}}{1-p_0^{3}},\\
&Pr\{A=(1~1~1)\}=\frac{(1-p_0)^{3}}{1-p_0^{3}}.
\end{align*}
A probability measure of $S$ is
\begin{align*}
&Pr\{S=(0~0)\}=\frac{3p_0^{2}(1-p_0)}{1-p_0^{3}},\\
&Pr\{S=(0~1)\}=\frac{2p_0(1-p_0)^{2}}{1-p_0^{3}},\\
&Pr\{S=(1~0)\}=\frac{p_0(1-p_0)^{2}}{1-p_0^{3}},\\
&Pr\{S=(1~1)\}=\frac{(1-p_0)^{3}}{1-p_0^{3}}.
\end{align*}
The probability measure of random vector $g'$ is
\begin{align*}
&Pr\{g'=(0)\}=p_0,\\
&Pr\{g'=(1)\}=1-p_0.
\end{align*}
From (\ref{equation2}),
\begin{align*}
p(1,3)=&p_0\cdot1\cdot p_0^{2}+(1-p_0)\Big(\frac{3p_0^{2}(1-p_0)}{1-p_0^{3}}\cdot p_0^{2}
+\frac{2p_0(1-p_0)^{2}}{1-p_0^{3}}\cdot p_0(1-p_0)\\
&+\frac{p_0(1-p_0)^{2}}{1-p_0^{3}}\cdot p_0(1-p_0)
+\frac{(1-p_0)^{3}}{1-p_0^{3}}\cdot(1-p_0)^{2}\Big)\\
=&p_0^{3}+\frac{3p_0^{4}(1-p_0)^{2}}{1-p_0^{3}}+\frac{3p_0^{2}(1-p_0)^{4}}{1-p_0^{3}}+\frac{(1-p_0)^{6}}{1-p_0^{3}}.
\end{align*}

(iii) $i=2$. Similarly, according to the conditional probability measure of $A$ and $S=A_{1}^{-1}A_2$, we can obtain a probability measure of $S$.
\begin{align*}
&Pr\{S=\begin{pmatrix}0 \\ 0\end{pmatrix}\}
=\frac{6p_0^{4}(1-p_0)^{2}+12p_0^{3}(1-p_0)^{3}}{-6p_0^{6}+24p_0^{5}-36p_0^{4}+30p_0^{3}-18p_0^{2}+6p_0},\\
&Pr\{S=\begin{pmatrix}0 \\ 1\end{pmatrix}\}
=\frac{2p_0^{3}(1-p_0)^{3}+2p_0^{2}(1-p_0)^{4}+2p_0(1-p_0)^{5}}{-6p_0^{6}+24p_0^{5}-36p_0^{4}+30p_0^{3}-18p_0^{2}+6p_0},\\
&Pr\{S=\begin{pmatrix}1 \\ 0\end{pmatrix}\}
=\frac{4p_0^{3}(1-p_0)^{3}+4p_0^{2}(1-p_0)^{4}+4p_0(1-p_0)^{5}}{-6p_0^{6}+24p_0^{5}-36p_0^{4}+30p_0^{3}-18p_0^{2}+6p_0},\\
&Pr\{S=\begin{pmatrix}1 \\ 1\end{pmatrix}\}
=\frac{6p_0^{2}(1-p_0)^{4}}{-6p_0^{6}+24p_0^{5}-36p_0^{4}+30p_0^{3}-18p_0^{2}+6p_0}.
\end{align*}
The probability measure of $g'$ is
\begin{align*}
&Pr\{g'=(0~0)\}=p_0^{2},\\
&Pr\{g'=(0~1)\}=p_0(1-p_0),\\
&Pr\{g'=(1~0)\}=p_0(1-p_0),\\
&Pr\{g'=(1~1)\}=(1-p_0)^{2}.
\end{align*}
Therefore,
\begin{align*}
p(2,3)=&p_0^{3}+\Big(12p_0^{6}(1-p_0)^{3}+36p_0^{5}(1-p_0)^{4}+24p_0^{4}(1-p_0)^{5}+36p_0^{3}(1-p_0)^{6}\\
&+12p_0^{2}(1-p_0)^{7}+6p_0(1-p_0)^{8}\Big)/\Big(-6p_0^{6}+24p_0^{5}-36p_0^{4}+30p_0^{3}-18p_0^{2}+6p_0\Big).
\end{align*}
To sum up,
\begin{align*}
P_{3\times 3}^{3}=&(1-p(0,3))(1-p(1,3))(1-p(2,3))\\
=&-24p_0^{9}+144p_0^{8}-360p_0^{7}+492p_0^{6}-414p_0^{5}+234p_0^{4}-90p_0^{3}+18p_0^{2}.
\end{align*}

It is easy to check that the number of full rank matrices of weight $0,1,2,3,4,5,6,7,8,9$ is equal to $0,0,0,6,36,72,36,18,0,0$, respectively.
Therefore,
\begin{align*}
P_{3\times 3}^{3}
=&6p_0^{6}(1-p_0)^{3}+36p_0^{5}(1-p_0)^{4}+72p_0^{4}(1-p_0)^{5}+36p_0^{3}(1-p_0)^{6}+18p_0^{2}(1-p_0)^{7}\\
=&-24p_0^{9}+144p_0^{8}-360p_0^{7}+492p_0^{6}-414p_0^{5}+234p_0^{4}-90p_0^{3}+18p_0^{2}.
\end{align*}
This shows that our derived formula is exact.

\section{Conclusion}\label{section6}
In this paper, we study the performance of SRLNC for reliable multicast and derive an exact expression for the decoding success probability of SRLNC. This is due to the fact that we propose a criterion that a vector is contained in a subspace. To obtain this criterion, we construct a basis of a subspace, with respect to this basis, the coordinates of a vector are known, based on a maximal linearly independent set of the columns of a matrix.
In addition, we show that the derived expressions presented in Th. \ref{theorem1} and Th. \ref{theorem2} can collapse to RLNC case, and the derived expression presented in Th. \ref{theorem1} can be scaled up to the well-known \textit{BKW} bound, and the derived expression presented in Th. \ref{theorem2} is equivalent to the existing expressions. The exactness and the computation of the derived expression are demonstrated by a simple example. Our future work targets at reducing the complexity of the derived expression to improve its utility.

\ifCLASSOPTIONcaptionsoff
  \newpage
\fi

% References
\bibliographystyle{IEEEtran}
\bibliography{myreferences}

\end{document}